\documentclass[twocolumn,showpacs,preprintnumbers,prb,fleqn,floatfix]{revtex4}

\usepackage{graphicx}
\usepackage{dcolumn}
\usepackage{bm}

\begin{document}
\title{{ In-plane current-voltage characteristics and oscillatory Josephson-vortex flow resistance in La-free
Bi$_{2+x}$Sr$_{2-x}$CuO$_{6+\delta }$ single crystals in high
magnetic fields}}

\author{S. I. Vedeneev$^{1,2}$ and D. K. Maude$^{1}$}

\affiliation{ $^1$ Grenoble High Magnetic Field Laboratory, Centre
National de la Recherche Scientifique, B.P. 166, F-38042 Grenoble
Cedex 9, France \\
$^2$ P.N. Lebedev Physical Institute, Russian Academy of Sciences,
119991 Moscow, Russia }

\date{\today }

\begin{abstract}
We have investigated the in-plane $I(V)$ characteristics and the
Josephson vortex flow resistance in high-quality La-free
Bi$_{2+x}$Sr$_{2-x}$CuO$_{6+\delta }$ (Bi2201) single crystals in
parallel and tilted magnetic fields at temperatures down to
$40$~mK. For parallel magnetic fields below the resistive upper
critical field $H^{*}_{c2}$, the $I(V)$ characteristic obey a
power-law with a smooth change with increasing magnetic-field of
the exponent from above 5 down to 1. In contrast to the
double-layer cuprate Bi2212, the observed smooth change suggests
that there is no change in the mechanism of dissipation (no
Kosterlitz-Thouless transition) over the range of temperatures
investigated. At small angles between the applied field and the
$ab$-plane, prominent current steps in the $I(V)$ characteristics
and periodic oscillations of Josephson-vortex flow resistance are
observed. While the current steps are periodic in the voltage at
constant fields, the voltage position of the steps, together with
the flux-flow voltage, increases nonlinearly with magnetic field.
The $ab$-flow resistance oscillates as a function of field with a
constant period over a wide range of magnetic fields and
temperatures. The current steps in the $I(V)$ characteristics and
the flow resistance oscillations can be linked to the motion of
Josephson vortices across layers.
\end{abstract}


\maketitle
\section{Introduction}

It is now well established that the stack of Josephson junctions
formed by the atomic layers of high-temperature superconductors
(HTSC) represents a nonlinear system with unique dynamic
properties. A magnetic field perpendicular to the layers induces
Abrikosov-type pancake vortices in the planes. In contrast, a
magnetic field applied along the CuO$_2$ layers (parallel to the
$ab$-planes) creates Josephson vortices, each carrying one flux
quantum, and whose cores are located between the superconducting
layers. When the parallel magnetic field exceeds the crossover
field, $H_{cr} = \Phi_0/\pi \gamma s^2$, where $\Phi_0$ is the
flux quanta, $\gamma$ is the anisotropy of the London penetration
depth, and $s$ is the interlayer spacing, the cores of Josephson
vortices start to overlap and a Josephson lattice is formed.
\cite{Bul73} At the same time, the vortex ground state and the
vortex phase diagram of highly-anisotropic (layered) HTSC in
parallel and oblique magnetic fields remains the open question.
The picture of a phase transition from a solid (phase-ordered)
state to a liquid state of a combined lattice structure of
Josephson vortices and pancake vortices in high magnetic fields
also remains unclear.\cite{Konc}

Due to the proximity to the Mott insulator, phase fluctuation are
strong and the temperature of the superconducting transition,
$T_c$, in cuprates may be governed by phase fluctuations. In two
dimensions, the phase fluctuations  can give rise to a
Kosterlitz-Thouless transition which is described by the thermal
unbinding of vortex-antivortex pairs. However, there is no clear
picture for the destruction of superconducting order in layered
HTSC via the unbinding of vortices. \cite{Lee}. Emery and Kivelson
\cite{Emery} have calculated the Kosterlitz-Thouless transition in
layered HTSC assuming that each layer is fluctuating
independently, even for systems with strongly Josephson coupled
bilayers. On the other hand, Corson \textit{et al.} \cite{Corson}
using microwave conductivity have confirmed the
Kosterlitz-Thouless nature of the phase transition but concluded
that in Bi$_2$Sr$_2$CaCu$_2$O$_{8+\delta}$ (Bi2212) system it is
the bilayer which should be considered as a unit, i.e., the
superconducting phase is strongly correlated between the two
layers of a bilayer. In view of these findings, it is interesting
to look for a Kosterlitz-Thouless transition in the single-layer
high-$T_c$ superconductor Bi$_{2}$Sr$_{2}$CuO$_{6+\delta }$ where
the interlayer spacing is considerable more. As previously
reported, \cite{Ando,Pradhan93,Iye} this can be achieved by
measuring the $I(V)$ characteristics of the samples. Bi2212 single
crystals, both in zero and in applied magnetic fields, show
non-Ohmic, power-law $I(V)$ characteristics. The observed
power-law behavior, $V\sim I^{\alpha(T,H)}$, with a characteristic
jump in the exponent $\alpha(T,H)$ near the superconducting
transition has been interpreted as evidence for the
Kosterlitz-Thouless transition.

It is apparent that to gain a better understanding of the vortex
ground state and vortex phase diagram in HTSC, it is essential to
understand the dynamics of Josephson vortices. In the generally
studied geometry, a $c$-axis external current drives Josephson
vortices in the direction perpendicular to both the current and
the magnetic field. The driven motion of the vortices is
responsible for the observed flow resistance. Koshelev
\cite{Koshelev00} calculated the flux-flow resistivity of the
Josephson vortex lattice in a layered superconductor and showed
that the magnetic field dependence of the flux-flow resistivity is
characterized by three distinct regions. At low magnetic fields
the flux-flow resistivity grows linearly with field. When the
Josephson vortices start to overlap the flux-flow resistivity
crosses over to a regime of quadratic magnetic field dependence.
Finally, at very high fields the flux-flow resistivity saturates
at a value close to the $c$-axis quasiparticle resistivity.

A Josephson lattice driven by a current along the $c$ axis,
together with the resulting flux-flow resistance, has been
intensively studied both experimentally and theoretically because
the coherent Josephson vortices motion gives rise to strong
resonant phenomena which may lead to very important applications
of high-$T_c$ superconductors. \cite{Zhu} In particular, the
flux-flow voltage creates an oscillating current through the
Josephson effect and the current excites Josephson plasma waves at
Terahertz frequency. Part of the energy may be emitted as
electromagnetic waves. \cite{Bul06} Under appropriate conditions,
coherent motion of Josephson vortices is possible and this can
excite cavity resonances of the stack (the Fiske resonances),
which in turn influence the vortex motion. Such geometric
resonances manifest themselves in the $c$-axis current voltage
($I(V)$) characteristic as a series of steps separated by a
constant voltage. \cite{Kleiner,Krasnov} Recently, Ooi \textit{et
al.} \cite{Ooi} reported novel periodic oscillations of the
flux-flow resistance with a $c$-axis bias current. The measured
periodicity, $H_p$, is given by the relation $\Phi_0/2ws$ where
$w$ is the sample size in the $ab$-plane along the direction
perpendicular to the applied magnetic field. Such oscillations of
the Josephson vortices flow resistance can be explained by a
matching effect between the lattice spacing of Josephson vortices
along the layers and the size of the sample.

It should be noted that in Josephson-coupled layered
superconductors, vortices exist for any orientation of the
magnetic field. In the presence of an electrical current, the
Lorentz force on vortices causes motion which generates both
in-plane and interplane electric fields and which induces a
flux-flow voltage drop across the superconducting sample. The
sliding motion of Josephson vortices along the $ab$-plane is
easily driven by a current along the $c$ axis. For a driving force
exerted along the $c$ axis (current in the $ab$-plane), the
CuO$_2$ planes effectively pin the Josephson vortices preventing
motion along the $c$ axis. \cite{Machida} The dynamics of the
Josephson lattice along the $c$ axis is far from being fully
understood. In tilted high fields $H>>H_{cr}$ at small angles
between the applied field and the $ab$-plane, a zigzag structure
along the $c$ axis arising from the attractive interaction of
pancake vortices with Josephson vortices, see Ref.
[\onlinecite{Bul96}]. In this case one might expect a series of
maxima in the plasma frequency and the critical current at angles
for which the pancake lattice is in resonance with the Josephson
lattice.

Investigations of the dynamics of the vortex lattice in HTSC are
usually limited to the region near $T_c$ because the large
critical current density $J_c$ makes low-temperature studies
difficult, whereas in clean systems the vortex lattice is expected
to melt due to thermal fluctuations. Values of the upper critical
field $H_{c2}$ and $J_c$ in the Bi$_{2}$Sr$_{2}$CuO$_{6+\delta}$
system are relatively low, which allows to observe the effect of
strong magnetic fields on the dynamics of the Josephson vortex
lattice at very low temperatures below and above a melting line
where the magnetic field behavior depends on the mechanism of
dissipation.

Here, we present an experimental study of the in-plane $I(V)$
characteristics and the in-plane Josephson vortex flow resistance
in parallel and tilted magnetic fields in single-layer La-free
Bi$_{2}$Sr$_{2}$CuO$_{6+\delta}$  single crystals at temperatures
down to $40$~mK. At small angles $\theta$ between the applied
field and $ab$-plane, we have found prominent current steps in the
$I(V)$ characteristics and periodic oscillations in the
Josephson-vortex flow resistance. To gain a better understanding
as to the nature of the observed current steps and periodic
oscillations, we have compared the behavior of the in-plane flux
flow resistance in Bi2201 single crystals with the published data
for Bi2212 with double CuO$_2$ layers.

\section{Experiment}

The high-quality  La-free Bi$_{2+x}$Sr$_{2-x}$CuO$_{6+\delta }$
(Bi2201) single crystals used for the present study were grown by
a KCl-solution-melt free growth method. \cite{Gorina,Martovitsky}
The preparation and characterization of Bi2201 single crystals are
described in detail elsewhere. \cite{Veden04} As before, in this
work we have used \textit{as-grown} single crystals. The
properties of the samples investigated in this work are summarized
in Table~\ref{table1}. The temperature dependence of the
resistance of the present samples in zero magnetic field are
essentially the same as those we reported previously.
\cite{Veden04} We have measured the resistance, $I(V)$
characteristics and the differential resistance $dV/dI(V)$
characteristics for Bi2201 single crystals using a standard
four-probe technique with symmetrical positions of the
low-resistance contacts on both $ab$ surfaces of the
sample.\cite{t1} The current was applied parallel to the
$ab$-plane in all cases. In the experimental arrangement used, the
crystal could be rotated $in-situ$ relative to the direction of
the magnetic field with an angular resolution better than
$0.1^{\circ}$. The $\theta=0^{\circ}$ orientation was precisely
determined from the lowest value of the resistance at a fixed
temperature. A configuration with $H\perp J$ was used always.
Measurement procedures in the resistive magnet are described in
detail in Refs. [\onlinecite{Veden04,Veden06}].

\begin{table}[h]
 \centering \caption{Doping ($p$), critical temperature
 ($T_{c}$) determined from the 50\% point of the resistive transition,
and dimensions for the Bi2201 single crystals investigated.
 Optimal doping in Bi2201 occurs at $p\simeq 0.17$ holes per
 Cu.}\label{table1}
 \begin{tabular}{c|c|c|c} \hline
 Sample & $p$~(holes/Cu) & $T_{c}$~(K) & Size~($l \times w \times d$)\\
 \hline
  \#5 & $0.17$ &  $9.8$ &  $1.8$~mm~$\times$~$0.7$~mm~$\times$~$2$~$\mu$m\\
  \#7 & $0.14$ &  $5.5$ &  $1.5$~mm~$\times$~$0.7$~mm~$\times$~$1$~$\mu$m\\
  \#4 & $0.13$ &  $3.6$ &  $3$~mm~$\times$~$0.3$~mm~$\times$~$1$~$\mu$m\\
  \#2 & $0.12$ &  $2.0$ &  $1$~mm~$\times$~$0.75$~mm~$\times$~$1$~$\mu$m\\
  \hline
 \end{tabular}
 \end{table}

\section{Results and discussion}
\subsection{Current-voltage characteristics}

As reported in Ref. [\onlinecite{Pradhan93}], a parallel external
magnetic field plays a crucial role in the dynamics of the
Kosterlitz-Thouless transition, and in particular, reduces the
transition temperature. In sufficiently strong magnetic fields
parallel to the $ab$-plane (up to 8~T), the Kosterlitz-Thouless
transition is suppressed because the magnetic field-induced
vortices increase the dissipation and reduces the stability of the
vortex-antivortex pairs. However, subsequently, it has been
suggested \cite{Hu} that in Bi2212 at high parallel magnetic
fields, there is a multi-critical point in the $H-T$ phase diagram
characterized by a magnetic field $H_{mc} = \Phi_0/2\sqrt{3}
\gamma s^2$, above which exists a novel intermediate
Kosterlitz-Thouless type phase. The existence of this intermediate
phase was also confirmed by the non-ohmic power-law $I(V)$
characteristics in the Bi2212 system. Chen and Hu \cite{Chen} very
recently investigated theoretically the in-plane $I(V)$
characteristics of interlayer Josephson vortices in cuprate HTSC
using a computer simulation and showed that for highly anisotropic
systems ($\gamma = 20$) and at high magnetic field, the power-law
in-plane $I(V)$ characteristics changes it's exponent from 1
(i.e., Ohmic) at high temperatures to $\ge 3$ at intermediate
temperatures, which can be attributed to a Kosterlitz-Thouless
transition.

In layered high-$T_c$ superconductors, the anisotropy is usually
expressed by the ratio $\gamma = \lambda_c / \lambda_{ab}$, where
$\lambda_c$ is the London penetration depth for super currents
along the $c$ axis, and $\lambda_{ab}$ is the penetration depth
for currents in the $ab$-plane. The anisotropy ratio $\gamma$ can
also be expressed by the ratio $ \sqrt{m_{c}/m_{ab}}$ between the
effective masses of the quasiparticles along the $c$-axis and in
the $ab$-plane, which can be related to the transport anisotropy
with $\gamma \simeq \sqrt{\rho _{c}/\rho _{ab}}$ using $\rho =
m/ne^{2}\tau $ for the out-of-plane resistivity $\rho _{c}$ and
in-plane resistivity $\rho _{ab}$. \cite{Kotliar,Schilling,Luo}.

Owing to the large value of $H_{c2}$ in parallel magnetic fields,
in order to study the power-law behavior of the $I(V)$
characteristics of Bi2201 single crystals, we have used underdoped
samples with a low $T_c$. Previously, we found \cite{Veden04} that
the normal-state anisotropy ratio $\rho_{c}/\rho _{ab}$ in high
magnetic fields and at low temperatures is practically temperature
independent. For this reason, we can safely neglect any variation
of $\gamma$ for the temperatures used here. For the underdoped \#7
($T_c=5.5$~K) and heavily underdoped \#2 ($T_c=2$~K) samples, the
anisotropy is equal to $\gamma \simeq 40$ and $\gamma \simeq 22$,
respectively. In this case the values of the multi-critical
magnetic field $H_{mc}$ should be $\simeq 10$~T and $\simeq 18$~T.

\begin{figure}
\includegraphics[width=1.0\linewidth,angle=0,clip]{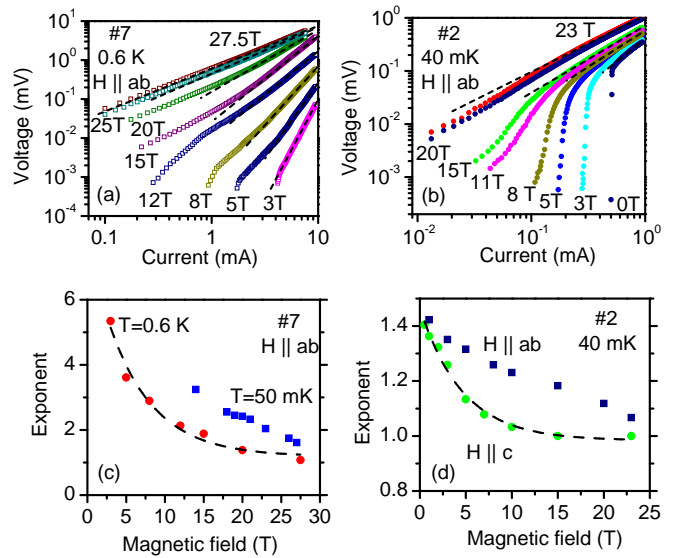}
\caption{\label{fig1} (color on-line) (a) $I(V)$ characteristics
plotted using a logarithmic scale for samples \#7 (a) and \#2 (b)
near the magnetic-field induced suppression of the
superconductivity at 0.6~K and 40~mK for selected magnetic fields
oriented parallel to the $ab$-plane. The dashed lines are
power-law fits to the data. (c) and (d) Magnetic-field dependence
of the power-law exponent for samples \#7 and \#2, respectively.}
\end{figure}

The in-plane current-voltage characteristics for samples \#7 and
\#2 near the magnetic-field induced suppression of
superconductivity at 0.6~K and 40~mK are plotted using a
logarithmic scale in Fig.~\ref{fig1}(a) and Fig.~\ref{fig1}(b),
for selected magnetic fields ($H\parallel ab$). The temperature
dependence of the resistance $R _{ab}(T)$ of the present samples
in zero magnetic field (not shown) are essentially the same as
those reported previously for samples with the same doping.
\cite{Veden04}

The $I(V)$ characteristics can be fitted to a power law for currents
above a critical current $I_c$ (dashed lines in
Fig.\ref{fig1}(a-b)). The extracted power law exponent, $\alpha$, of
the $I(V)$ characteristics, is plotted in Figures~\ref{fig1}(c) and
\ref{fig1}(d) as a function of magnetic field for samples \#7 and
\#2, respectively, at 0.6~K, 50~mK and 40~mK. At $T=0.6$~K, $\alpha$
decreases rapidly with increasing field and its value tends to unity
in high magnetic fields, while at very low temperature, $\alpha$
decreases much more slowly. However, although the $I(V)$
characteristics show a power-law behavior with the change in the
value of exponent from 1 to above 5 [Fig.~\ref{fig1}(c)] and the
magnetic fields far exceed the $H_{mc}$ value, both the figures show
that $\alpha$ decreases smoothly with increasing field, without any
features that might be interpreted as evidence for a transition to
the Kosterlitz-Thouless type phase. It should be noted that the
magnetic fields 23~T and 28~T applied here are far in excess of the
fields at which the 3D vortex melting occurs in low $T_c$ Bi2201
samples. \cite{Morr}

It has been suggested in Refs. [\onlinecite{Pradhan92,Ando}], that
the power law dependence of the $I(V)$ curves at low temperatures
is related to the interplay of the in-plane vortex lines and the
two-dimension pancake vortices created by a small component of the
magnetic field parallel to the $c$ axis. That can appear because
of a small misalignment to the $c$ axis within the crystal.
However, we also observe a power-law behavior in the $ab$-plane
$I(V)$ characteristics when the magnetic field is applied
perpendicular to $ab$-plane. In Fig.~\ref{fig1}(d), we also plot
the power-law exponent versus magnetic field applied along the
$c$-axis for crystal \#2 at $T=40$~mK. One can see that $\alpha$
lies in the same range for both orientations of the field. In view
of this, it is difficult to imagine that a possible small $c$-axis
component of the magnetic field (in the parallel geometry) gives
the same change in the value of $\alpha$ as the large magnetic
field oriented along the $c$-axis. In addition, we can formally
exclude macroscopic sample inhomogeneity, as the origin of the
observed the power-law behavior in $I(V)$ characteristics in
Fig.~\ref{fig1}(a) and Fig.~\ref{fig1}(b), the crystals are of a
very high quality judging from the small rocking curve width $\le
0.1^{\circ }$.

\subsection{Differential resistance}

In high magnetic fields, the in-plane $I(V)$ characteristics in
low bias region (small currents) showed unusual behavior for a
superconductor. This feature in the $I(V)$ characteristic in the
vicinity of the zero bias can be more clearly seen the measured
first derivatives $dV/dI(V)$ for the sample \#2 with current and
voltage contacts on the same surface of the crystal at a
temperature of $T=40$~mK with the magnetic field oriented
perpendicular and parallel to the $ab$-plane.

\begin{figure}
\includegraphics[width=1.0\linewidth,angle=0,clip]{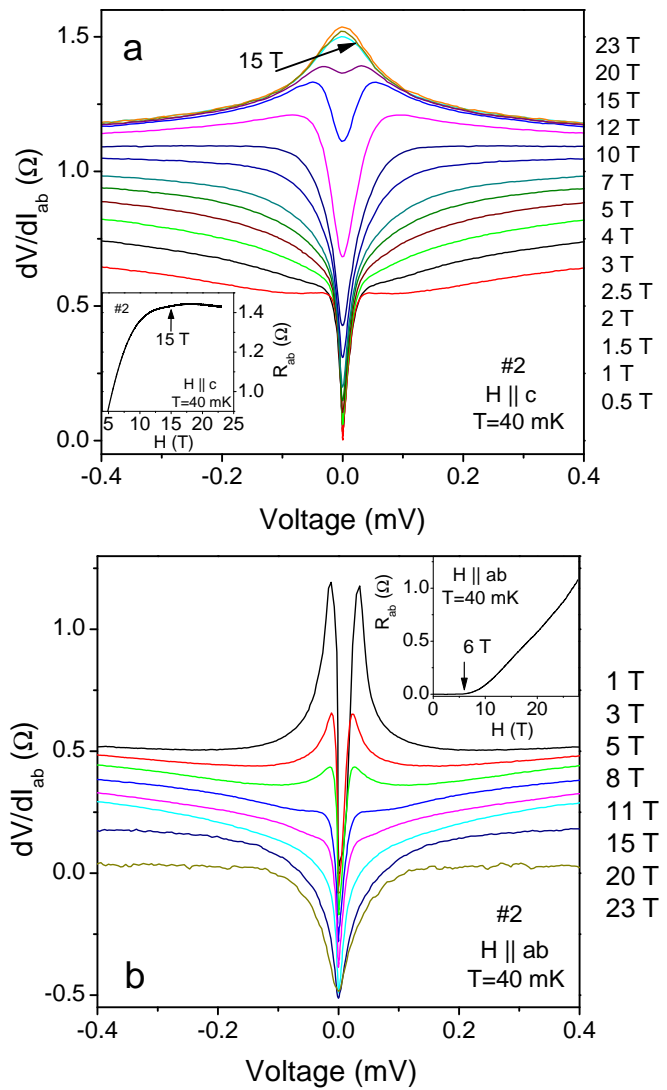}
\caption{\label{fig2} (color on-line) Differential resistance
$dV/dI(V)$ near the current induced suppression of superconductivity
at $T=40$~mK for different magnetic fields applied along the
$c$-axis of the crystal \#2 (a) and along $ab$-plane (b). For
clarity, the curves in (b) have been shifted vertically (downwards)
with respect to the upper curve. The inset in (a) shows the
magnetoresistance $R_{ab}(H)$ measured at $T=40$~mK for the sample
\#2 with the magnetic field applied along the $c$-axis. The inset in
(b) shows the magnetoresistance $R_{ab}(H)$ measured at $T=40$ mK
for the sample \#2 in a parallel magnetic field ($H\parallel ab$).}
\end{figure}

In Fig.~\ref{fig2}(a) (main panel), we plot the voltage dependence
of the differential resistance $dV/dI(V)$ near the current induced
suppression of the superconductivity in different magnetic fields
applied along the $c$-axis of the crystal \#2 at $T=40$~mK. In the
inset, we plot the the magnetoresistance $R_{ab}(H)$ measured at
$T=40$~mK for the same sample with the magnetic field applied
along the $c$-axis. As can be seen, the differential resistance
$dV/dI(V)$ saturates at high fields in accordance with the
exponent $\alpha(H)$ in Fig.~\ref{fig1}(c) and Fig.~\ref{fig1}(d).
The $dV/dI(V)$ curves have a minimum close to zero bias voltage,
the amplitude of which decreases with increasing magnetic field.
This minima should disappear at a magnetic field corresponding to
the transition of the sample to the normal state. What is observed
is slightly different since this minimum transforms to a maximum
for magnetic fields $\geq 15$~T after the suppression of the
superconductivity (marked by arrows in the inset and main panel).
It can be supposed that such a behavior of the differential
resistance in the vicinity of low bias at magnetic fields above
the resistive upper critical field $H^{*}_{c2}$ is connected with
moving vortex-like excitations which have been previously observed
in superconducting cuprates at $T> T_c$ in magnetic fields by the
detection of a Nernst signal \cite{Wang} and measurement of an
angular dependence of the magnetoresistance on
Y$_{1-x}$Pr$_x$Ba$_2$Cu$_3$O$_{7-\delta}$ single crystals.
\cite{Sandu} With increasing current, the number of vortices
decreases and the differential resistance in Fig.~\ref{fig2}(a)
decreases rapidly.

Figure~\ref{fig2}(b) (main panel) shows the analogous $dV/dI(V)$
curves for the same sample in the magnetic fields applied in the
$ab$-plane of the crystal at $T=40$ mK. For clarity, the curves
have been shifted vertically with respect to the upper curve. In
the inset we show the magnetoresistance $R_{ab}(H)$ measured at
$T=40$ mK for the same sample in a parallel magnetic field (the
onset of the superconducting transition is marked by an arrow).
Despite the different types of vortex lattice (the pancake
vortices in perpendicular and Josephson vortices in parallel
configurations), the $dV/dI(V)$ curves in Fig.~\ref{fig2}(b) have
a general form identical to that in Fig.~\ref{fig2}(a). At our
maximum field $H=28$~T, the resistance of the sample has only
reached 70\% of its normal-state value and the minimum of the
differential resistance in the vicinity of low bias currents is
not completely suppressed. This suggests that the mechanism of the
dissipation in the neighborhood of $H^{*}_{c2}$ is common to both
magnetic field directions.

These results allows us to conclude that the behavior of the flux
flow resistance in the $ab$-plane in Bi2201 single crystals at low
temperatures is largely similar to that in Bi2212 except for the
absence of a transition to the Kosterlitz-Thouless phase.
Apparently, this transition does not take place at low
temperatures even in high magnetic fields. Most likely the
Kosterlitz-Thouless transition in Bi2201 can be observed only in a
very narrow temperature region near $T_c$.

\subsection{Current steps in I(V) characteristics}

The interaction of the pancake vortices with the Josephson
vortices can lead to features in the $I(V)$ characteristics with
the transport current along the $ab$-plane when the period of the
pancake vortex lattice matches the period of the Josephson vortex
lattice. In order to rule out the presence of spatial
inhomogeneity, which can be present in underdoped samples and
which may influence the results found, we have used for these
measurements an optimally doped Bi2201 single crystal (\#5) with
$T_c=9.8$~K. To investigate the influence of misalignment between
the magnetic field and the superconducting layers, the sample was
rotated at small angles around its position of parallel to the
layers. Temperatures for these measurements were chosen so that
the available magnetic field would suffice to observe the
flux-flow voltage while remaining below the melting line. With
increasing current in the $ab$-plane, we have found prominent
current steps in the $I(V)$ characteristics. The amplitude of the
current steps depends on the angle between the applied field and
the $ab$-plane. At very small angles $\theta$, as in the parallel
configuration of field, no current steps are observed in the
$I(V)$ characteristics. This is consistent with the sample
remaining free of pancake vortices until the normal component
$H\sin\theta$ exceeds the perpendicular critical field $H_{c1}$
for pancake formation. \cite{ Maslov} For a detailed
investigation, we chose the orientation of the field for which the
amplitude of the current steps was a maximum.

\begin{figure}
\includegraphics[width=0.9\linewidth,angle=0,clip]{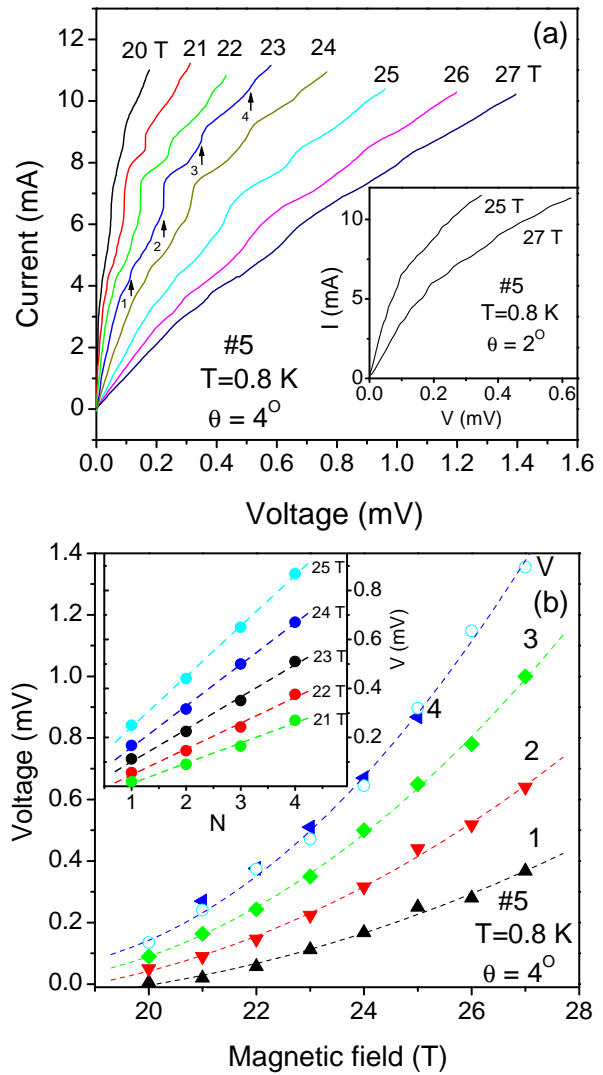}
\caption{\label{fig3} (color on-line) (a) $I(V)$ characteristics
of sample \#5 with $T_c=9.8$~K in various tilted ($\theta =
4^{\circ}$) magnetic fields at 0.8~K with the transport current
flowing in the $ab$-plane perpendicular to the field. Current
steps are indicated by arrows. The inset in Fig.~\ref{fig3}(a)
shows similar characteristics for two different magnetic fields at
$\theta=2^{\circ }$. (b) Voltage position of the current steps
versus magnetic field (solid symbols) The number of the current
step is indicated for each curve. Open circles show the flux-flow
voltage for a current of $10$~mA. The dashed lines are fits to the
experimental data. The inset shows the voltage position of the
current steps versus the step number for different fields.}
\end{figure}

Figure~\ref{fig3}(a) (main panel) displays the $I(V)$
characteristics for the single crystal \#5 in various tilted
magnetic fields at 0.8~K with transport current flowing in the
$ab$-plane perpendicular to the applied magnetic field. Current
steps are clearly observed (for $H=23$~T, they are marked by
arrows). The angle $\theta$ between the applied field and the
$ab$-plane was equal to $4^{\circ}$. The inset in
Fig.~\ref{fig3}(a) shows similar characteristics for two different
magnetic fields at $\theta=2^{\circ }$.

In Fig.~\ref{fig3}(b) (main panel), we plot the voltage position
of the current steps versus magnetic field (the number of the
current step are indicated for each curve). It can be seen that
the voltage position of each step increases nonlinearly with
magnetic field (solid symbols) as well as the flux-flow voltage,
at the current 10~mA showed by open circles. It is evident that
the data can be fitted quite well to curves of the form
$V=aH+bH^2$ where $a$ and $b$ are constants (dashed lines).

These steps cannot be attributed to Fiske steps caused by an
interaction  of the ac Josephson effect with electromagnetic modes
of a cavity formed by the Josephson junctions (geometric
resonance) for a number of reasons. Firstly, the dc voltage is
applied along the $ab$-plane instead of across the junctions and
secondly, the voltage position of the current steps depends on the
magnetic field which is not expected for Fiske resonances.
Furthermore, while the current steps are periodic in the voltage
at constant fields, the step period increases with increasing
magnetic field. This can be seen in the inset of
Fig.~\ref{fig3}(b) where we show the voltage position of the
current steps versus the step number for different fields.

On the other hand, the range of the magnetic fields in
Fig.~\ref{fig3}(a) at $\theta = 4^{\circ}$ exceeds the crossover
field $H_{cr}$ for sample \#5 and also the field component
perpendicular to the layers exceeds the perpendicular lower
critical field $H_{c1}$ for pancake formation. Under these
conditions, according to the model of Bulaevskii \textit{et al.}
\cite{Bul96}, the tilted magnetic field induces a triangular
lattice of Josephson vortices and pancake vortices which form a
zigzag structure along the $c$ axis. In this case one might expect
a series of maxima in the $c$-axis plasma frequency and the
$c$-axis critical current at angles where the pancake lattice is
in resonance with the Josephson lattice. Such zigzag structure
result in oscillations of the out-of-plane resistance and a sharp
increase of the $c$-axis critical current in Bi2212 single crystal
at 30~K in field 20~T at small angles near $\theta=0^{\circ}$ with
width $\approx 2^{\circ}$. \cite{Bul96} Based on the model of
Bulaevskii\textit{et al.} \cite{Bul96}, we estimate that the
periods of the Josephson and pancake lattices in our experiment,
for example, at $H=23$~T should be matched at $\theta \simeq
1^{\circ}$. This is in disagreement with the observed maximum
amplitude of the current steps at $\theta \simeq 4^{\circ}$. We do
not understand the reason for this discrepancy. However,
subsequent studies \cite{Tsui, Kakeya} after Ref.
[\onlinecite{Bul96}] have shown that the maximum of such resonance
in Bi2212 single crystals at $T=5$~K takes place at the angle
$\theta=5^{\circ}$ and the angular dependence of the resonance
field is much more drastic than the result of Ref.
[\onlinecite{Bul96}].

In our case, with the magnetic field component parallel to the
layers and the current flowing in the $ab$-plane, the Lorentz
force acting on the vortices drives them parallel to the $c$ axis
with jumps across the superconducting layers. The motion of the
Josephson lattice generates both in-plane and out-of-plane
alternating electric fields and currents, which are coupled to
electromagnetic plasma waves. There should exist a strong
resonance emission when the velocity of the lattice matches the
velocity of the plasma wave. \cite{Koshelev01} The matching of the
Josephson frequency with the frequency of the emission may
manifest itself as the current steps not only in the $c$-axis
current, but possibly in the in-plane current as well
[Fig.~\ref{fig3}(a)]. Since the behavior of the plasma mode in
parallel and tilted fields is poorly understood at present, in
contrast to the perpendicular magnetic field geometry, we are
unable to discus here this topic more fully. However, since it was
assumed that the periodic steps in the in-plane $I(V)$
characteristics may be related to the moving Josephson vortex
lattice along the $c$ axis, we have tried to experimentally
establish this fact.

\subsection{Flow resistance oscillations}

As noted above, Ooi \textit{et al.} \cite{Ooi} found periodic
oscillations of the $c$-axis flow resistance of Josephson vortices
as a function of the in-plane magnetic field in small-strip Bi2212
single crystal in a wide range of temperatures and fields. A
resistance of the strip sample in used geometry was almost equal
to the $c$-axis resistance of the intrinsic junctions and the
contribution of in-plane resistivity was neglected. These
oscillations were related to a matching effect between the lattice
spacing of Josephson vortices along the layers and the width of
the sample. Subsequently, Koshelev \cite{Koshelev01} and Machida
\cite{Machida} have reproduced the oscillation by calculating the
flux flow resistance by taking the surface barrier effect into
consideration.

\begin{figure}
\includegraphics[width=0.9\linewidth,angle=0,clip]{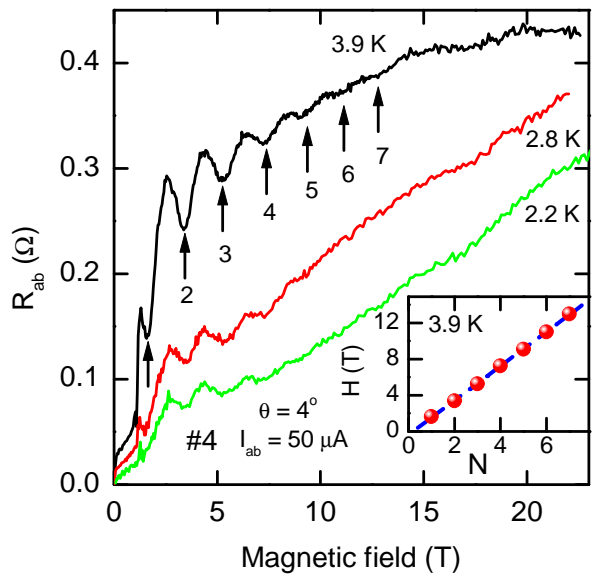}
\caption{\label{fig4} (color on-line) Flow resistance of the
Josephson vortices as a function of magnetic field for sample \#4
with an applied current of $50 \mu A$ at different temperatures.
The inset shows the magnetic field position of the resistance
minima (marked by arrows in the main panel) versus the number
($N=1$ corresponds to the minima closest to zero bias) at
$T=3.9$~K.}
\end{figure}

An attempt to observe the periodic oscillations of the $ab$-plane
flow resistance of Josephson vortices in a parallel magnetic field
similar to Ref. [\onlinecite{ Ooi}] in our crystal with usual
sizes \cite{Veden04} has been unsuccessful. Because of this, we
have measured the in-plane Josephson vortex flow resistance versus
parallel and tilted magnetic fields in a Bi2201 strip single
crystal \#4 with $T_c=3.6$~K. In spite of the low $T_c$, a careful
characterization of the crystal showed the high quality, the high
homogeneity and the structural perfection of the sample. As the
magnetic field was oriented parallel to the $ab$-planes exactly,
the $ab$-flow resistance increased smoothly with increasing field
because only the number of vortices increased. At the same time,
the resistance begins to oscillate as a function of the magnetic
field for any small deviation of the field from the exactly
parallel geometry as shown by three curves in Fig.~\ref{fig4}. The
amplitude of oscillations gradually diminishes with increasing
magnetic field and oscillations are not observed at fields above
15~T. The oscillations also rapidly vanish when the angle $\theta$
between the applied field and the $ab$-plane exceeded $\approx
6^{\circ}$. The maximum amplitude of the oscillations was reached
at $\theta \approx 4^{\circ}$.At the same time, the oscillations
are most prominent at higher temperatures where thermal
fluctuations reduce the role of background pinning and the crystal
behaves as if it contains fewer defects than it actually does.

It can be seen from the inset in Fig.~\ref{fig4} that the period
of the oscillations $\Delta H \approx 1.9$~T is to a good
approximation constant over a wide range of magnetic fields. The
period also does not depend on the temperature (see curves for
three different temperatures in Fig.~\ref{fig4}). The oscillations
can be observed for almost all temperatures below $T_c$, but have
reduced amplitude at very low temperatures and near $T_c$.

The flow resistance oscillations in Fig.~\ref{fig4} are quite
similar to the oscillations of the $c$-axis Josephson-vortex flow
resistance in Bi2212 single crystals previously observed by Ooi
\textit{et al.} \cite{Ooi} . The oscillations in
Ref.[\onlinecite{Ooi}] originate from a matching between the
triangular lattice and the width $w$ of the junction perpendicular
to the magnetic field direction. The period of the oscillations,
$H_p$, corresponded to adding one flux quantum $\Phi_0$ per two
layers in the crystalline lattice. The minima in the resistance
corresponded to the condition where the magnetic field was $nH_p
\approx nH_0/2$ ($n$ is an integer). Here $H_0=\Phi_0/ws$ is the
field needed to add one vortex per layer in a Bi2212, where $s$ is
the periodicity of the CuO$_2$ double layers.

In Ref. [\onlinecite{Ooi}], a transport current flowing across the
layers exerts a Lorentz force on the Josephson vortices in the
plane of layers. An important parameter is the critical current
above which the lattice starts to move, producing a finite
voltage. This current is determined either by bulk pinning or by
interaction with the boundaries. \cite{Koshelev02}

For the measured flow resistance in Fig.~\ref{fig4}, a transport
current flowing along the layers increases the effect of the
Lorentz force on the Josephson vortices across the layers. Due to
the interaction with the boundaries of the sample, the period of
oscillations in Fig.~\ref{fig4} should depend on the size of the
sample along the $c$-axis direction. The flow resistance shows
periodic oscillations with period $\Delta H \approx 1.9$~T which
is very close (within the accuracy of measuring the sample
thickness) to the estimated $H_0=\Phi/ds=1.7$~T, corresponding to
the magnetic field needed to add one vortex quantum per layer in
the crystalline lattice. Here $d$ is the size of our strip single
crystal along $c$ axis (thickness) and $s=12.3$~\AA is the
distance between the superconducting CuO$_2$ layers in Bi2201.
This result suggests that Josephson vortices in Bi2201 form a
square lattice in the ground state. As before, the oscillations
are most prominent in the high temperature regime where thermal
fluctuations reduce the role of background pinning and the crystal
behaves as if it contained fewer defects than it actually does.

We note that somewhat similar commensurate oscillations of the
$ab$-plane resistivity for magnetic fields nearly aligned with the
$ab$-plane has been observed on YBa$_2$Cu$_3$O$_{7-\delta}$ single
crystals in earlier work by Gordeev \textit{et
al.}\cite{Gordeev00}. These oscillations are due to the
oscillatory melting line separating the vortex liquid from the
vortex smectic state and they are periodic in $H^{-1/2}$. This is
very different from the periodic in $H$ oscillations of the
vortex-flow resistance reported in Ref.[\onlinecite{Ooi}] and
observed here, which in addition are observed over a wide range of
temperatures and magnetic fields below and above a melting line.

\subsection{Discussion}

Finally, it is pertinent to recall here a transport phenomenon
which has its origins in a Lorentz-force-independent dissipation
that has been observed in strongly coupled layered Bi- and
Tl-based materials.\cite{Iye,Woo,Ando,Pradhan93} For this case the
resistivity in the non-Ohmic regime did not depend on the angle
between the magnetic field and current when they are both parallel
to the layers. The problem of an orientation-independent
dissipation and the role of the Lorentz force in  the motion of
the Josephson vortex lattice cannot be explained by simple
flux-flow theory and is far from being totally understood.

Turning back to Fig.~\ref{fig3}(a), we can suppose that the motion
of the Josephson lattice generates a traveling electromagnetic
wave with the $ab$-plane wave vector selected by the lattice
structure. One might expect resonance phenomena when the velocity
of the lattice matches the plasma wave velocity. \cite{Koshelev01}
On the other hand, if the motion of the Josephson vortex lattice
generates a voltage across the layers, one would expect that the
current steps in Fig.~\ref{fig3}(a) are due to the matching of the
Josephson frequency with the frequency of the plasma mode. Both
mechanisms may result in the current steps in the $I(V)$
characteristics.

\section{conclusion}

In summary, we have studied the in-plane $I(V)$ characteristics
and Josephson vortex flow resistance in La-free Bi2201 single
crystals in parallel and tilted magnetic fields at temperatures
down to $40$~mK. For parallel magnetic fields, we find a power-law
$I(V)$ characteristic with a smooth decrease with magnetic-field
of the exponent describing the power-law. At small angles between
the applied field and $ab$-plane, prominent current steps in the
$I(V)$ characteristics and periodic oscillations of
Josephson-vortex flow resistance are observed. While the current
steps are periodic in the voltage at constant fields, the voltage
position of the steps, together with the flux-flow voltage,
increases nonlinearly with increasing magnetic field. The
$ab$-flow resistance oscillates as a function of magnetic field
with a constant period over a wide range of magnetic fields and
temperatures. The current steps in the $I(V)$ characteristics and
the flow resistance oscillations can be linked to the motion of
the Josephson vortices across layers.

\begin{acknowledgments}
This work has been partially supported by PICS grant No. 3447. One
of us (S.I.V.) was partially supported by the Russian Foundation
for Basic Research Projects No. 06-02-22001  and No. 07-02-00349.
The work at GHMFL was partially supported by the European 6th
Framework Program under contract number RITA-CT-3003-505474.
\end{acknowledgments}


\begin{thebibliography}{37}
\bibitem[{t1()}]{t1}
\bibinfo{note}{The contacts on both $ab$ surfaces of the sample also allows us
  to measure $\rho_{c}$ which is required to determine the transport
  anisotropy.}
\end{thebibliography}

\end{document}